\documentclass{template}
\usepackage{hyperref}
\usepackage{multicol}

\begin{document}

\catchline{1}{1}{2019}{}{}
\markboth{M. S. Feser \& I. Krumphals}{Preparing pre-service physics teachers to diagnose students’ conceptions not covered by physics education textbooks}

\title{Preparing pre-service physics teachers to diagnose students’ conceptions not covered by physics education textbooks}

\author{Markus Sebastian Feser}

\address{Physics Education, Universität Hamburg\\
Hamburg, Free and Hanseatic City of Hamburg, Germany\\
\email{markus.sebastian.feser@uni-hamburg.de}}

\author{Ingrid Krumphals}
\address{Institute for Secondary Teacher Education, University College of Teacher Education Styria\\
Graz, Styria, Austria\\
\email{ingrid.krumphals@phst.at}}

\maketitle


\begin{abstract}
To date, there is a lack of research on learning environments for pre-service physics teachers that allow them to learn and practise diagnosing students’ conceptions that are (currently) not covered by physics education textbooks (e.g. students' conceptions about viscosity). In this study, we developed and piloted such a learning environment, which was implemented and piloted twice in a seminar for pre-service physics teachers. As coping with a diagnostic process is particularly demanding for pre-service physics teachers, our accompanying research aims to identify learning barriers within our developed learning environment. The results indicate that the participants experience the learning environment with varying degrees of difficulty. One main difficulty for pre-service physics teachers seems to be in interconnecting their content knowledge with their pedagogical content knowledge in the diagnostic process.
\end{abstract}


\begin{multicols}{2}
\section{Introduction}


Diagnostic activities play a crucial role in the physics classroom to foster students' learning processes. For example, physics teachers implement tasks during instruction that address students’ prior knowledge and design learning environments sensitive to and that reveal the needs of the students. Consequently, teachers need profound assessment literacy.\cite{6} Specifically, they need to be able to identify students’ perspectives and their challenges and difficulties during instruction in order to adopt teaching properly.\cite{7,8} 

\begin{figure*}
	\begin{center}
		\includegraphics[width=6in]{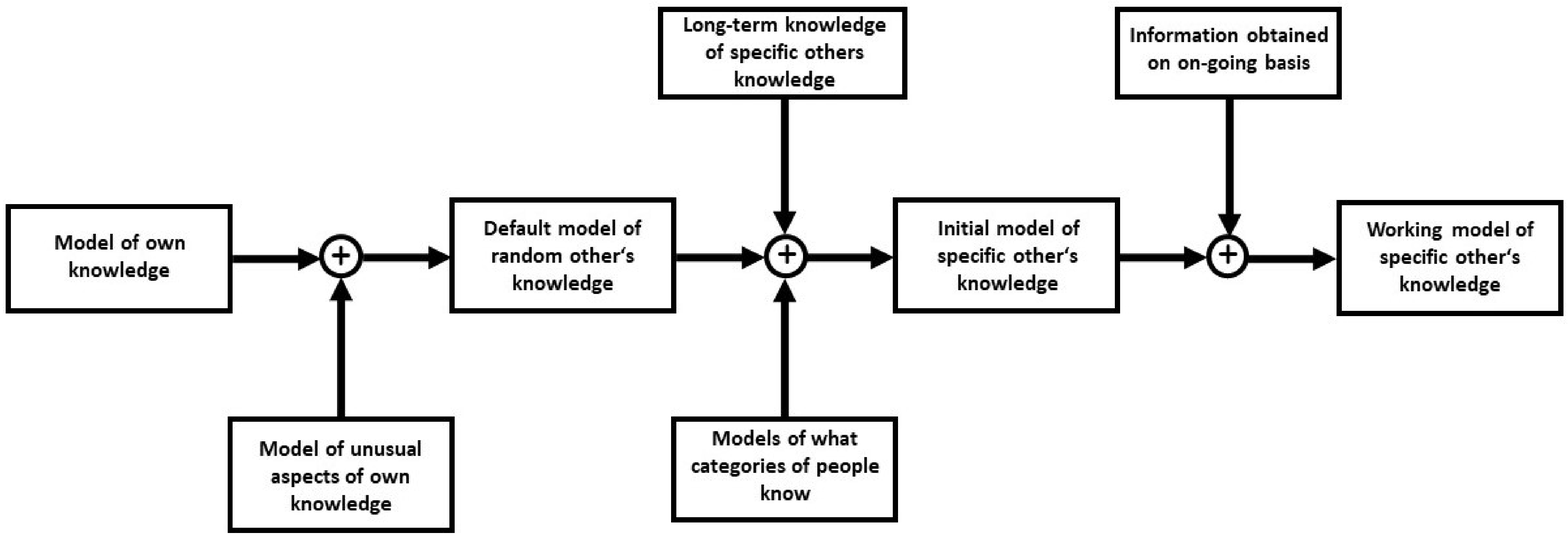}
	\end{center}
	\caption{Nickerson’s model of one’s working model of specific others’ knowledge.}
	\label{fig1}
\end{figure*}

Therefore, profound diagnostic skills of teachers regarding students’ conceptions\footnote{In this paper, the term ‘conception’ refers to underlying patterns in students’ explanations of scientific facts and phenomena (for more details on how the term ‘conception’ is defined in physics education research see Ref.~\refcite{19,17}).} are imperative for adaptive and high-quality physics teaching.\cite{1} Many typical students’ conceptions are well described in physics education textbooks (e.g. Ref.~\refcite{2,18}). Such textbook descriptions are essential for physics teachers to design learning environments suited to meet the needs of their students. However, there is also a variety of students’ conceptions encountered by physics teachers in their day-to-day teaching that are (currently) not covered by physics education textbooks, for example, conceptions about subsidiary or new topics within school physics. As a result, to properly support students’ learning progress in the physics classroom, teachers also require knowledge on how to manage ‘off-textbook’ students’ conceptions. Therefore, diagnosing both textbook and off-textbook students’ conceptions is a daily, essential and demanding task for physics teachers.

During teacher education, pre-service physics teachers should be adequately prepared for this task. To date, however, there is a lack of research on learning environments for pre-service teachers that allow them to learn and practise diagnosing off-textbook students’ conceptions. In this exploratory study, we developed and piloted a learning environment that focuses on the diagnosis of students’ conceptions about the viscous behaviour of fluids—an example for off-textbook students’ conceptions (see section \ref{sec2}). As the diagnosis of students’ conceptions is challenging for pre-service physics teachers,\cite{4} our accompanying research focuses on identifying elements within our developed learning environment that promote or hinder pre-service teachers’ learning processes in terms of the acquisition of diagnostic skills to refine the learning environment by following a design-based research approach.\cite{12,5} 

In this paper, we describe the structure of our developed learning environment. Moreover, we report the initial findings of our accompanying research regarding the first implementation of the developed learning environment.

\section{Theoretical and general background of the study}\label{sec1}
Put simply, the goal of diagnosing students’ conceptions in the physics classroom is to gain and evaluate information on students’ knowledge and thoughts. Therefore, it is important to conceptualise how such information-gathering and evaluation processes generally work. Suitable for this purpose is Nickerson’s\cite{9} heuristic model of one’s working model of specific others’ knowledge (see Figure \ref{fig1}). In this model, it is assumed that the evaluating person (in our case, the physics teacher) uses their own knowledge as an anchor for the knowledge of the person being evaluated (in our case, the students). In a first step, the evaluating person makes an adjustment to this anchor by excluding the facets of their own knowledge that they perceive as unusual or exceptional. Through this, the evaluating person derives a default model of a random other’s knowledge. In the second step, this default model is further adjusted into an initial model regarding the knowledge of the specific person whose knowledge is to be evaluated. This is done by directly considering already existing knowledge about the evaluated person and by inferring the knowledge of the evaluated person based on their (presumed) group membership (e.g. being a 10th-grade secondary school student). Finally, in the third step, this initial model is continuously modified by integrating the information the evaluating person receives through an interaction with the person being evaluated to form an increasingly refined working model of the evaluated person’s knowledge.

Consequently, teachers must pass through a complex diagnostic process to validly diagnose students’ conceptions. Ideally, teachers plan their diagnostic actions, execute these actions to collect adequate data to perform a profound diagnosis and then reflect on their actions and diagnosis. To map this complex process of diagnosing students’ conceptions, Krumphals and Haagen-Schützenhöfer developed a diagnostic process model.\cite{10,11} In this model, the process in which teachers diagnose students’ conceptions in the physics classroom is ideally conceptualised, and it consists of three steps that can be repeated sequentially: (1) pre-actional phase, (2) actional phase and (3) post-actional phase (for more details see Ref.~\refcite{11}). Based on this diagnostic process model an intervention for pre-service physics teachers (5th-6th semester of bachelor studies) was developed at the University of Graz to enhance their diagnostic skills related to students’ conceptions.\cite{11} This intervention (four 2-hour units in total) is currently implemented and is an integral part of the physics teacher education program for the south-eastern region of Austria. In terms of its learning objectives, this intervention focuses on students’ conceptions, which can be easily found in physics education textbooks. 

Teachers need to manage students’ conceptions that are well known in physics education research and that are comprehensively covered in physics education textbooks (e.g. Ref.~\refcite{2,18}). However, in the physics classroom, teachers are often confronted with students’ conceptions that they have never heard of during their studies or have not read about in any textbook. Moreover, teachers are also confronted with students’ statements that indicate new and unfamiliar conceptions (e.g. conceptions regarding subsidiary or new topics in school physics) and that also need to be considered to appropriately adapt instruction and promote individual students’ learning. Within such scenarios, physics teachers cannot draw on textbook knowledge about students’ conceptions. Instead, the physics teachers must infer students’ conceptions themselves, for example, by interpreting and evaluating their students' statements based on their physics content knowldege, their pedagogical content knowledge and/or their practical school knowlege.\cite{20} To summarize, in day-to-day physics teaching, teachers need to be able to do both: diagnose textbook as well as off-textbook students’ conceptions. Consequently, teacher education should prepare pre-service physics teachers to meet this requirement.

\section{Learning environment on how to diagnose off-textbook students’ conceptions}\label{sec2}
To address the requirement for physics teachers to diagnose both textbook and off-textbook students’ conceptions properly in initial teacher training, we extended the original intervention on textbook students’ conceptions (see section \ref{sec1}; Ref.~\refcite{11}) by adding a learning environment covering an additional 2.5-hour unit. Initially, we planned to design this unit as a face-to-face session, but due to the COVID-19 pandemic, the unit needed to be rearranged as an online session to meet the physical distancing requirements.

In terms of its content and learning objectives, the unit focuses on diagnosing students’ conceptions not represented in physics education textbooks on an exemplary level (in our case, students’ conceptions of the viscous behaviour of fluids; for details see Ref.~\refcite{14,3,15,16}). To this end, the participants worked in small groups during the unit and practised diagnosing off-textbook students’ conceptions, which were evident in video-recorded and authentic statements of two secondary school students (Jonas and Tim; see transcript excerpts\footnote{For illustrative purposes, all transcript excerpts in this paper have been edited, pseudonymized, and approximately translated.} 1 and 2). Specifically, the statements of the secondary school students addressed an everyday  phenomenon in which the viscous behaviour of a fluid can be observed: dropping a spoon into a jar of honey. \\

\noindent {\bf Transcript excerpt 1:}

{\bf Interviewer:} ‘Can you explain why the spoon drops slowly in honey but not in water?’

{\bf Jonas:}  ‘I do not know, maybe... Maybe because the honey is a little sticky. The contained sugar is sticky.’\\

\noindent {\bf Transcript excerpt 2:}

{\bf Interviewer:} ‘Can you explain the reason for the spoon slowly sinking in honey compared to a faster sinking in water?’

{\bf Tim:}  ‘It is more work done by the spoon. […] You need more strength to displace the honey than to displace the water.’

{\bf Interviewer:} ‘And why is it that you need more strength to displace honey?’

{\bf Tim:}  ‘I would say because the density of honey is higher than the density of water.’\\

\begin{figure*}
	\begin{center}
		\includegraphics[width=4.5in]{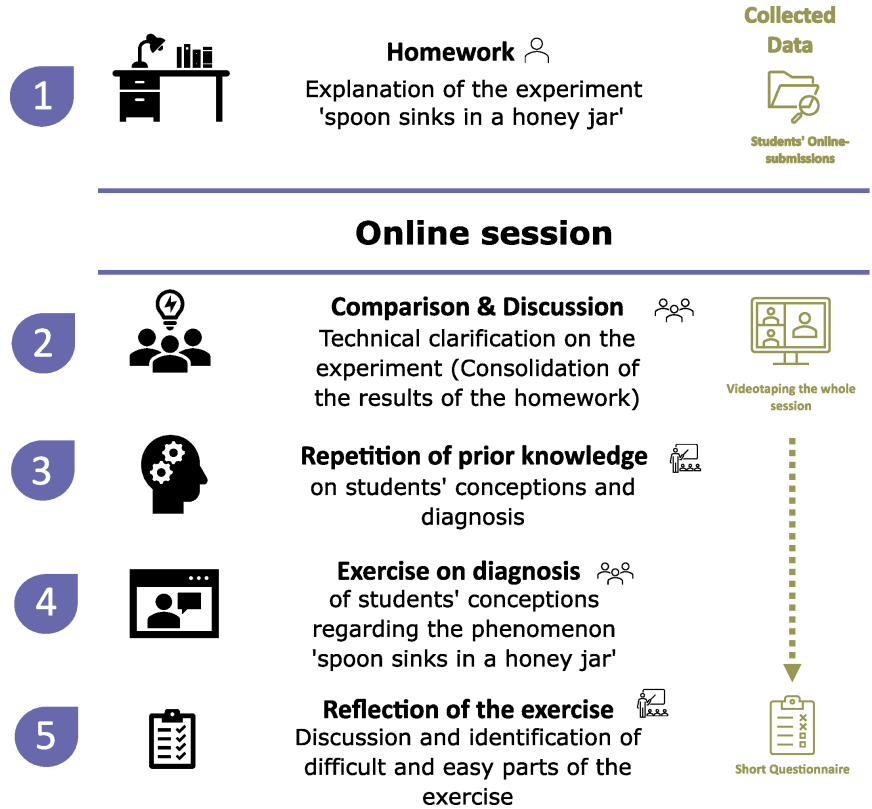}
	\end{center}
	\caption{Design of the learning environment and data collection.}
	\label{fig2}
\end{figure*}

As illustrated in Figure \ref{fig2}, the developed learning environment is structured into five phases. Phase 1 comprises preliminary homework that the pre-service physics teachers are requested to work on autonomously and submit prior to the unit. In this homework, the participants were asked to explain in their own words why a spoon slowly sinks when dropped into a jar of honey and to compare their first explanation with the explanation found in scientific literature (a selection of suitable university-level textbook excerpts was provided to them for this comparison task). This homework aimed to establish the basis for a joint clarification of this phenomenon at the beginning of the unit from a scientific point of view. In addition, the comparison of one’s own explanation to the explanation found in the literature intends to promote pre-service physics teachers’ sensitivity to the complexity and interconnectedness of everyday explanations and those from the scientific literature.

After a brief introduction to the setting of the online session in phase 2, the pre-service physics teachers were asked to discuss the results of their homework in small groups to compare and consolidate their results (approximately 30 minutes). Subsequently, an inter-group summary and, as far as necessary, further clarification took place in a plenary session (approximately 15 minutes). 

Phase 3 (approximately 15 minutes) began with a multiple-choice task on the diagnosis of students’ conceptions, which aimed to cognitively activate the participants’ prior knowledge. The pre-service physics teachers first worked on this task individually. Their aggregated answers were then compared and formatively evaluated, and possible further questions were clarified in the plenary. This was followed by an impulse talk and a plenary discussion on the possible strategies to diagnose students’ conceptions not covered by textbooks on physics teaching. Finally, phase 3 concludes with a central message to be sensitive to what students say, especially when diagnosing students’ conceptions that are not found in physics education textbooks.

Phase 4 (approximately 45 minutes) involves a group exercise of diagnosing off-textbook students’ conceptions. The main goal of this phase is to provide pre-service physics teachers with an authentic situation in which they can practise and reflect on how to diagnose off-textbook students’ conceptions. To this end, the participants were asked to diagnose students’ conceptions using video vignettes of authentic students’ statements. As mentioned above, these authentic video vignettes showed Jonas and Tim reporting their observations on a spoon dropping and slowly sinking into a jar of honey and providing their individual explanations for this observation (see transcript excerpts 1 and 2). During this task, since students' conceptions are underlying patterns reflected in students' statements, the pre-service physics teachers could not evaluate students’ conceptions directly (precisely, that  Tim confused the viscosity of honey with its density, whereas Jonas confused viscosity and stickiness). They must infer these conceptions by decoding and interpreting the corresponding students’ statements, a requirement that can be challenging for pre-service physics teachers. Therefore, at the end of phase 4, the participants were requested to share and discuss which parts of the diagnostic process they perceived as manageable or challenging. 

Finally, phase 5 (approximately 30 minutes) aims to add to the previous discussion of the pre-service physics teachers in their groups. They were asked to reflect on the previous exercise and share their thoughts and insights with all participants in the plenary. To structure and consolidate the results of this collaborative reflection, the participants’ most important insights on how to deal with off-textbook students’ conceptions in the physics classroom were collected and discursively scripted in a shared document.

\begin{figure*}
	\begin{center}
		\includegraphics[width=3.47in]{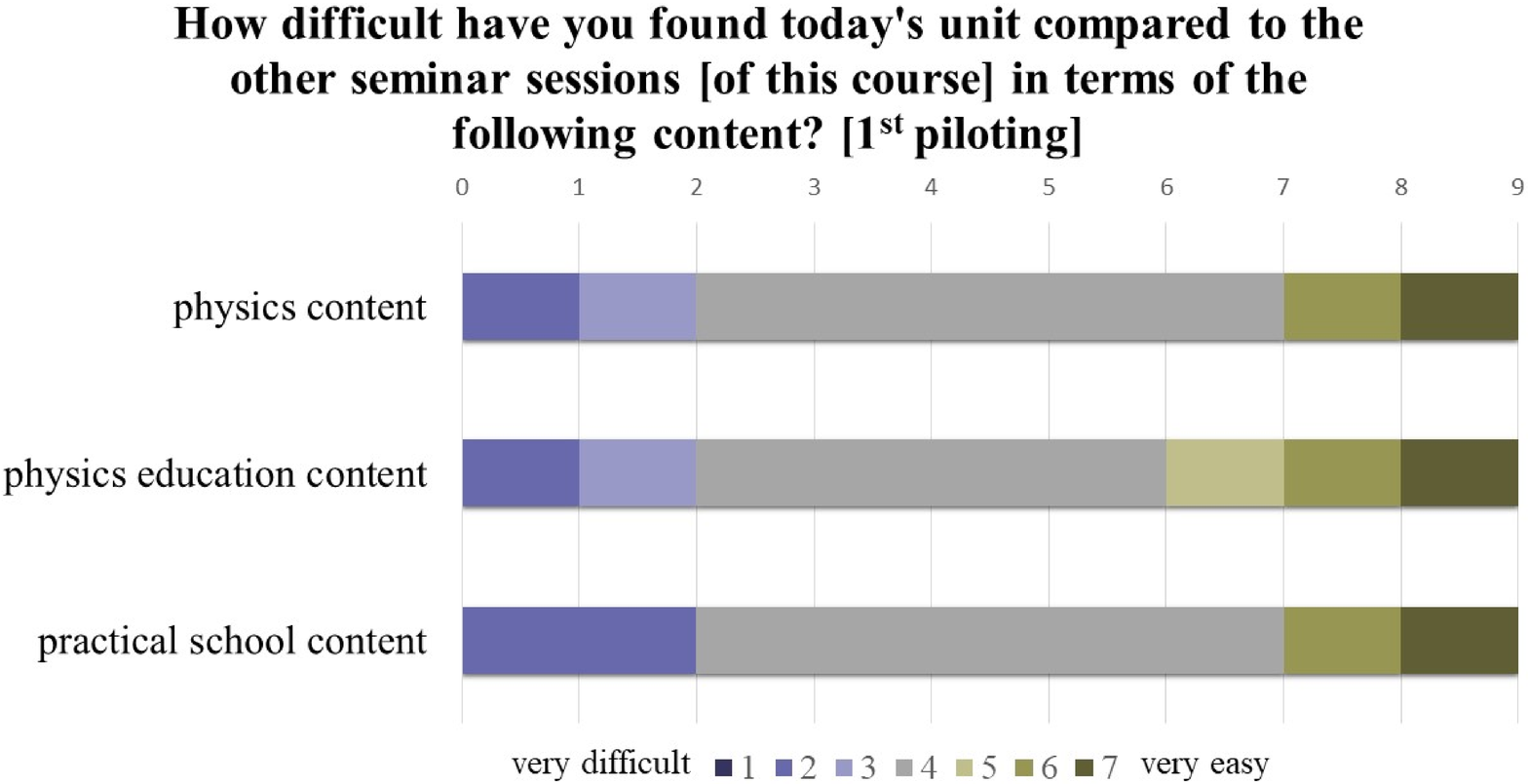}	\includegraphics[width=3.47in]{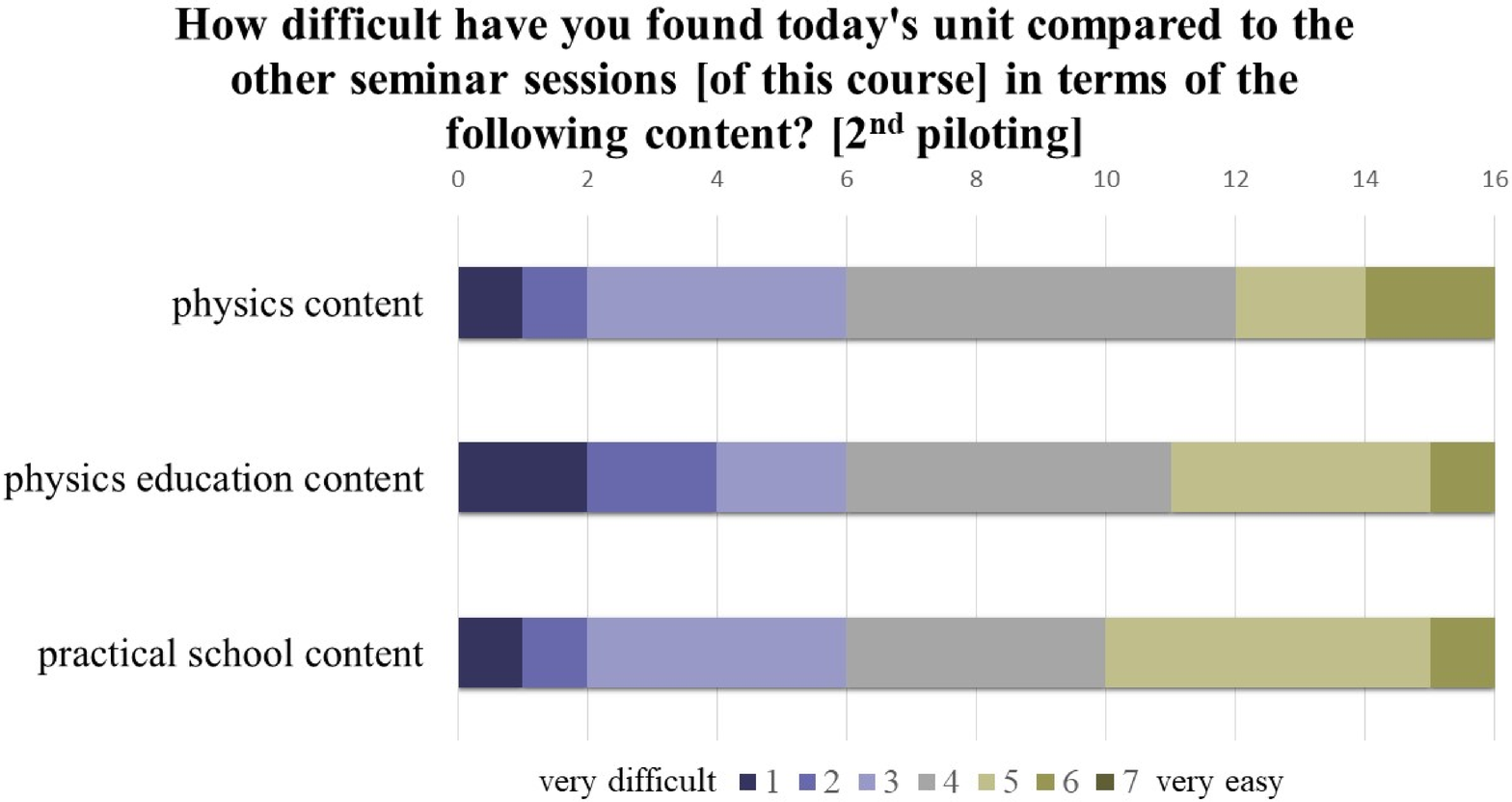}
	\end{center}
	\caption{Perceived difficulty by pre-service physics teachers in several domains during the learning environment (first piloting and second piloting).}
	\label{fig3}
\end{figure*}

\section{Data collection for the accompanying research}
In May and December 2021, we piloted the learning environment in an online team-teaching unit with pre-service physics teachers (May 2021: N = 11, sixth semester; December 2021: N = 16, fifth semester) at the University College of Teacher Education Styria (Austria) who had already completed the original intervention on textbook students’ conceptions (see section \ref{sec1}; Ref.~\refcite{11}). The main goal of our accompanying research was to identify the learning barriers and elements that support pre-service physics teachers in their learning process related to diagnosing off-textbook students’ conceptions. To collect valid empirical data for the accompanying research, all sessions, including all group activities, were videotaped (see Figure \ref{fig2}). In addition, we used a short questionnaire at the end of the session to obtain initial feedback from the pre-service physics teachers on the elements of the learning environment they perceived as (less) beneficial and/or helpful. The feedback from the short questionnaire served as a guiding reference for our qualitative analysis of the video data.

\section{Initial results of the accompanying research} 

The feedback from the short questionnaire revealed that the participants expressed heterogeneous views regarding the difficulty of the learning environment’s content (see Figure \ref{fig3}). During the pilot implementation, some participants perceived the physics content, the physics education content and/or the practical school content as particularly challenging, while others perceived them as quite manageable. This suggests that the developed learning environment contains elements that both hinder and promote the participants’ learning. Thus, to refine our learning environment accordingly, the primary focus of the qualitative analysis of the video data, particularly the transcribed recordings of the group activities, is to identify and characterise the learning barriers and promoting elements in depth. 

Currently, this qualitative analysis is not yet completed and is still ongoing. However, our preliminary findings have already revealed some remarkable insights, especially into the challenges pre-service physics teachers face when practising diagnosing off-textbook students’ conceptions:

 The transcripts of phase 4 of the learning environment revealed that most of the participants found it manageable to detect whether the statements of a secondary school student contained something erroneous (see transcript excerpts 3 and 4) but found it difficult to connect this notion with their own content knowledge. Furthermore, inferring students’ conceptions by decoding and interpreting their statements seemed challenging for some pre-service physics teachers. Thus, Michael and Lisa emphasised that it was easier for them to connect with secondary school students’ thoughts if they possessed a profound understanding of a phenomenon and that it was beneficial to elementarise this understanding in advance (see transcript excerpt 4). However, most pre-service physics teachers were very vague about their interpretations and seemed to be uncertain whether their interpretations were valid or not. For example, Paul and Eric highlighted that secondary school students could just lack subject-specific vocabulary and that this could be the reason why they could not explain the phenomenon adequately (see transcript excerpt 3). \\

\noindent {\bf Transcript excerpt 3:}

{\bf John:} ‘So, what did you find easy or difficult, respectively? […] It was easy to identify the wrong statements. But it was kind of hard to match that to a student’s conception.’

{\bf Eric:}  ‘Yeah, adequately matching a statement to a core idea. I think that’s difficult. You can tell what’s wrong. But finding out what the underlying idea that leads to this false statement is not so easy.’

{\bf Mary:}  ‘We actually recognised the idea of the students quite quickly. Reducing that to a core idea was not so easy.’

{\bf Eric:} ‘I think it’s also difficult to know whether it’s really that core idea or if we only interpreted the statements in this way. […] We can’t tell whether that’s really their conception or just our interpretation.’

{\bf Paul:} ‘Yeah, sure! For example, the student might just have expressed it clumsily, and we inferred a student’s conception from this clumsy expression.’

{\bf Eric:} ‘Exactly! It could really just be because of the physics vocabulary that he couldn’t express adequately what he wanted to say. And we think that he doesn’t understand the concept.’\\

\noindent {\bf Transcript excerpt 4:} 

{\bf Susan:} ‘Identifying incorrect statements was easy.’

{\bf Michael:} ‘Exactly, but then phrasing or sorting them out was difficult (laughs). And what you also need in any case is that you clarify the phenomenon from a physics point of view beforehand. […] And then elementarise the whole thing in advance so that you know what the student is actually aiming at.’

{\bf Lisa:} ‘It’s definitely necessary to know the scientific explanation of a phenomenon. But I think you should also work out the basic ideas behind them in advance.’

{\bf Michael:} ‘Yes. Do you all agree?’\\

Overall, during the discussion and reflection activity in phases 4 and 5 of the learning environment, the participants predominantly referred to the aspects of the diagnostic process that they perceived as challenging rather than those that they perceived as manageable. The difficulties that were addressed frequently by the pre-service physics teachers were as follows: they perceived the learning environment’s content (especially the physics content and the physics education content) as fragmented rather than interlinked, and they found it too demanding to connect their content knowledge with their pedagogical content knowledge when practising off-textbook diagnosing students’ conceptions. Only in exceptional cases (e.g. transcript excerpt 4) were the participants able to build on synergies between their content knowledge and their pedagogical content knowledge. Therefore, it may be reasonable to assume that a major barrier in our learning environment is that it does not offer enough opportunities for all participants to defragment and interconnect the physics content and the physics education content taught to them.

\section{Discussion and outlook} 

From the initial results of our accompanying research, some pre-service physics teachers struggled to draw on their content knowledge when practising to diagnose off-textbook students’ conceptions. However, following Nickerson’s\cite{9} model, one’s own content knowledge is an essential prerequisite for evaluating another person’s knowledge and, therefore, students’ conceptions. If pre-service physics teachers possess a sound base of content knowledge, they can activate their own knowledge before following a diagnostic process. Our learning environment has some elements covering this aspect (specifically phases 1 and 2), but strengthening these elements is important.

In addition, our initial results indicate that, in accordance with Nickerson’s\cite{9} model, pre-service physics teachers should also draw on pedagogical content knowledge when practising diagnosing off-textbook students’ conceptions. However, pre-service physics teachers find it very demanding to connect their content knowledge with their pedagogical content knowledge during this process. Consequently, they struggle to interpret secondary school students’ statements and infer them to underlying conceptions. This result also emerged in other studies regarding the diagnostic competencies of pre-service physics teachers (e.g. Ref.~\refcite{11,13}). Therefore, it is plausible to assume that this finding represents a common difficulty of pre-service physics teachers when practicing to diagnose students' conceptions. Accordingly, future physics education research is necessary to investigate the generalizability of this finding. 

Within our developed learning environment, one way to address the challenge of pre-service physics teachers to infer underlying conceptions from students' statements might be to provide them with assistive learning materials when working with the video vignettes (e.g. guidelines to help them to interpret students' statements appropriately). However, what these learning materials should look like is not clear at this point, as the analysis of our collected video data is still in progress. Further in-depth results are needed to address this particular question.

\section*{Acknowledgments}
We thank the Department of International Affairs of the Universität Hamburg (Germany) for funding this study.

\end{multicols}
\noindent\rule[1ex]{\textwidth}{1pt}

\paragraph{Dr. Markus Sebastian Feser} works as a postdoc at the Department of Education of the Universität Hamburg in Germany. Before working at the Universität Hamburg, he studied education sciences, physics and mathematics at the Julius-Maximilians-Universität Würzburg in Germany. In 2019, he completed his PhD in education studies with his thesis on the role of language in physics teachers’ everyday assessment practice. His main research focus is the professional development of (pre-service) science teachers. Among other topics, his current research focuses on the sense of belonging of students and student teachers to science, the conceptions of school students of the viscous behavior of fluids, and the role of language in teaching and learning physics.

\paragraph{Prof. Dr. Ingrid Krumphals} has been a professor of physics education at the University of Teacher Education Styria in Austria since December 2020. Previously, she was a post-doctoral research fellow at the University of Graz and there her research focus was the development of diagnostic competence of preservice physics teachers. Ingrid Krumphals earned her PhD at the Austrian Educational Competence Centre of Physics at the University of Vienna. She also has 5 years of experience as a high school teacher in physics and mathematics. Currently, her research focuses on content-specific teaching and learning processes in physics teacher education and training as well as in physics education at secondary level.


\begin{thebibliography}{9}

\bibitem{6} S. K. Abell and M. A. Siegel, Assessment Literacy: What Science Teachers Need to Know and Be Able to Do, {\it The Professional Knowledge Base of Science Teaching}, eds. D. Corrigan, J. Dillon and R. Gunstone (Springer, Dordrecht,, 2011), pp. 205–221.

\bibitem{7} M. Brunner, Y. Anders, A. Hachfeld and S. Krauss, The Diagnostic Skills of Mathematics Teachers, {\it Cognitive Activation in the Mathematics Classroom and Professional Competence of Teachers}, eds. M. Kunter, J. Baumert, W. Blum, U. Klusman, S. Krauss and M. Neubrand (Springer, Boston, 2013), pp. 229–248.

\bibitem{8} R. Duit, H. Gropengießer, U. Kattmann, M. Komorek and I. Parchmann, The Model of Educational Reconstruction – a framework for improving teaching and learning science, {\it Science education research and practice in Europe: Retrospective and prospective}, eds. D. Jorde and J. Dillon (Sense Publishers, Rotterdam, 2012), pp. 13–27.

\bibitem{1} F. Vogt and M. Rogalla, Developing Adaptive Teaching Competency through coaching, {\it Teaching and Teacher Education} {\bf 25}, 1051–1060 (2009).

\bibitem{19} H. Niedderer and H. Schecker, Towards an explicit description of cognitive systems for research in physics learning, {\it Research in Physics Learning—Theoretical Issues and 
Empirical Studies}, eds. R. Duit, H. Goldberg and H. Niedderer(Leibniz-Institut für die Pädagogik der Naturwissenschaften und Mathematik, Kiel, 1992), pp. 74–98.

\bibitem{17} T. Plotz, I. Krumphals and C. Haagen-Schützenhöfer, Delphi study on the term ‘students’ conceptions’. {\it Journal of Physics: Conference Series} {\bf 1929}, 012006 (2021). 

\bibitem{2} R. Driver, E. Guesne and A. Tiberghien, {\it Children’s Ideas in Science} (Open University Press, Buckingham, 1985).

\bibitem{18} R. Stavy and D. Tirosh, {\it How students (mis-)un- derstand science and mathematics: intuitive 
rules} (Teachers College Press, New York, 2000).

\bibitem{4} J.-R. Wang, H.-L. Kao and S.-W. Lin, Preservice teachers' initial conceptions about assessment of science learning: The coherence with their views of learning science, {\it Teaching and Teacher Education} {\bf 26}, 522–529 (2010).

\bibitem{12} S. Barab and K. Squire, Design-Based Research: Putting a Stake in the Ground, {\it Journal of the Learning Sciences} {\bf 13}, 1–14 (2004).

\bibitem{5} C. Haagen-Schützenhöfer and M. Hopf, Design-based research as a model for systematic curriculum development: The example of a curriculum for introductory optics, {\it Physical Review Physics Education Research} {\bf 16}, 020152 (2020).

\bibitem{9} R. S. Nickerson, How we know—and sometimes misjudge—what others know: Imputing one's own knowledge to others, {\it Psychological Bulletin} {\bf 125}, 737–759 (1999).

\bibitem{10} J. Klug, S. Bruder, A. Kelava, C. Spiel and B. Schmitz, Diagnostic competence of teachers: A process model that accounts for diagnosing learning behavior tested by means of a case scenario, {\it Teaching and Teacher Education} {\bf 30}, 38–46 (2013).

\bibitem{11} I. Krumphals and C. Haagen-Schützenhöfer, Development of a Learning Environment to Enhance Preservice Physics Teachers’ Diagnostic Competence in Terms of Students’ Conceptions, {\it EURASIA Journal of Mathematics, Science and Technology Education} {\bf 17}, em1972 (2021).

\bibitem{20} H. E. Fischer and A. Kauertz, Professional competencies for teaching physics, {\it Physics Education}, eds. H. E. Fischer and R. Girwidz (Springer, Cham, 2021), pp. 25–53).

\bibitem{14} P. Eastwell, Bernoulli? Perhaps, but What About Viscosity?, {\it The Science Education Review} {\bf 6}, 1–13 (2007).

\bibitem{3} L. Faltin and M. S. Feser, Secondary school students’ conceptions about the viscous behaviour of liquids, {\it  Physics Education} {\bf 56}, 35018 (2021).

\bibitem{15} M. S. Feser and S. Mangal, Exploring preschoolers' conceptions about the viscosity of honey, {\it  STEM Education} {\bf 2}, (2022).

\bibitem{16} J. Medová, Z. Sedmáková, B. Uhrecký and L. Valovičová, Designing activities to develop statistical literacy in primary pupils while conducting physics laboratory work in informal settings,  {\it } {\bf 12}, 246 (2022).

\bibitem{13} A. C. Alonzo and C. von Aufschnaiter C, Moving Beyond Misconceptions: Learning Progressions as a Lens for Seeing Progress in Student Thinking, {\it The Physics Teacher} {\bf 56}, 470–473 (2018).


\end{thebibliography}
\end{document}